\documentclass{aa}  

\usepackage{soul}
\usepackage{ulem}
\usepackage{cancel}
\usepackage{graphicx}	
\usepackage{amsmath}	
\usepackage{color}
\usepackage{natbib}
\usepackage{url}
\usepackage{placeins}
\usepackage{xfrac}
\usepackage{comment}
\usepackage[breaklinks=true]{hyperref} 
\usepackage{amssymb}
\usepackage{orcidlink}
\usepackage{verbatim}
\usepackage{tablefootnote}

\defcitealias{Hunt2024}{HR24}
\defcitealias{Hunt2023}{HR23}

\definecolor{tali}{RGB}{184,20,124}

\definecolor{vale}{rgb}{0.8,0,0}

\DeclareRobustCommand{\VAN}[3]{#2}
\let\VANthebibliography\thebibliography
\def\thebibliography{\DeclareRobustCommand{\VAN}[3]{##3}\VANthebibliography}

\usepackage{txfonts}
\begin{document}

   \title{Binary and Grouped Open Clusters: A New Catalogue}


   \author{Palma, T.\inst{1,2}\orcidlink{0000-0002-0732-2737}
          \and
          Coenda, V. \inst{1,3}\orcidlink{0000-0001-5262-3822} 
          \and
          Baume, G.\inst{4,5}\orcidlink{0000-0002-0114-0502}
          \and
          Feinstein, C.\inst{4,5}\orcidlink{0000-0003-2341-0494}
          }

     \institute
   {
    Observatorio Astronómico, Universidad Nacional de Córdoba, Laprida 854, X5000BGR, Córdoba, Argentina
    \and
    Consejo Nacional de Investigaciones Cient\'ificas y T\'ecnicas de la Rep\'ublica Argentina (CONICET)
    \and
    Instituto de Astronom\'ia Te\'orica y Experimental (IATE), CONICET, Universidad Nacional de C\'ordoba, Laprida 854, X5000BGR, C\'ordoba, Argentina
    \and
    Facultad de Ciencias Astronómicas y Geofísicas, Universidad Nacional de La Plata, Argentina	
    \and
    Instituto de Astrofísica de La Plata (IALP), CONICET, Universidad Nacional de La Plata, Argentina	
\\
    \email{tpalma@unc.edu.ar}
   }

   \date{Received ---, 2024; accepted ---, 2024}

 
  \abstract
   {Understanding the formation and evolution of star clusters in the Milky Way requires precise identification of clusters that form binary or multiple systems. Such systems offer valuable insight into the dynamical processes and interactions that influence cluster evolution.
}
   {This study aims to identify and classify star clusters in the Milky Way as part of double or multiple systems. Specifically, we seek to detect clusters that form gravitationally bound pairs or groups of clusters and distinguish between different types of interactions based on their physical properties and spatial distributions.}
   {We used the extensive star cluster database of Hunt \& Reffert (2023, 2024), which includes 7167 clusters. By estimating the tidal forces acting on each cluster through the tidal factor ({\it TF}), and considering only close neighbours (within 50 pc), we identified a total of 2170 star clusters forming part of double and multiple systems. Pairs were classified as Binaries (B), Capture pairs (C), or Optical pairs (O/Oa) based on proper motion distributions, cluster ages, and color-magnitude diagrams.}
   {Our analysis identified 617 paired systems, which were successfully classified using our scheme. Additionally, we found 261 groups of star clusters, each with three or more members, further supporting the presence of multiple systems within the Milky Way that exhibit significant tidal interactions.}
   {The method presented provides an improved approach for identifying star clusters that share the same spatial volume and experience notable tidal interactions. }
   \keywords{open clusters and associations: general -- star clusters: general -- Catalogs -- Astrometry}

   \maketitle
  
%
\section{Introduction}

\begin{figure*}
\centering
\includegraphics[trim=0 0 0 0, width=0.9\textwidth]{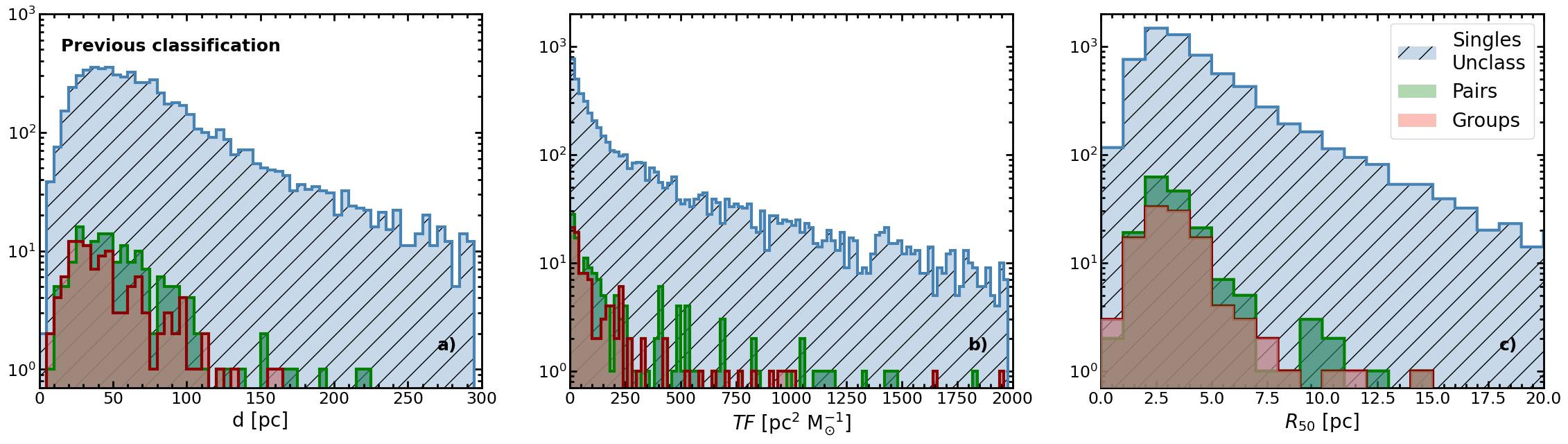}\\
\caption{Distribution of distances, computed tidal factors ({\it TF}), and sizes of cataloged star clusters considering different classifications.}
\label{fig:dist}
\end{figure*}
\begin{figure*}
\centering
\includegraphics[trim=0 0 0 0, width=0.9\textwidth]{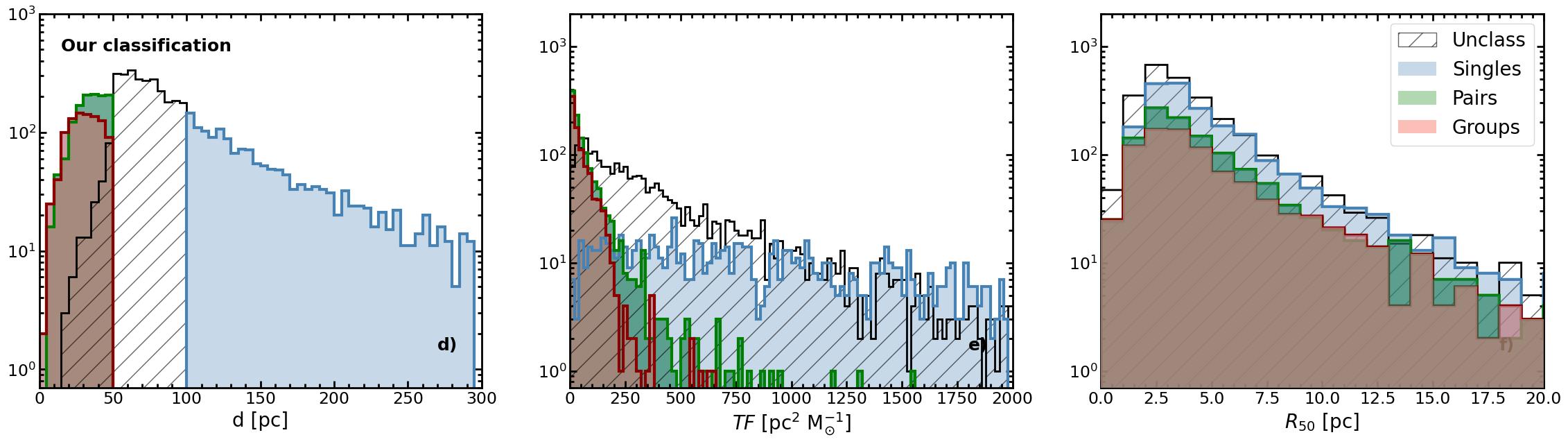}\\
\caption{Same as Fig.\ref{fig:dist} but based our classification criteria (see text).}
\label{fig:dist2}
\end{figure*}

Star clusters are fundamental tools for studying the formation and evolution histories of the galaxy to which they belong. In particular, star clusters forming binary (or multiple) systems are of great interest on a global scale, constituting about 10-15\% of the total cluster population \citep{Priyatikanto19, Casado21a}. These clumps become part of a system linked by different pathways. In some cases, the components of the system have been formed at the same time of the same primordial cloud \citep{Priyatikanto16}. Consequently, they share similar characteristics such as age, chemical composition, and distance. 
In other cases, sequential formation may occur when the stellar evolution of one of the clusters induces the collapse of a nearby cloud, either by stellar winds or supernova shocks, and thus generates the formation of a companion cluster \citep{ Goodwin97}.
Another possibility is the dynamical capture of a cluster through tidal forces \citep{vandenBergh96}  or resonant capture \citep{delaFuente09}. In this case the clusters form independently and may have significantly different properties.

Detection of multiple systems has been an area of interest since the 1990s, with significant advancements in the past decade. This effort has led to the development of several catalogs, each covering specific subsets of clusters based on particular selection criteria
\citep[e.g.,][among others]{delaFuente09, Soubiran19, PieckaPaunzen21, Casado21b, Song22}. Table  \ref{tab:ref} provides a comprehensive compilation of 
these published catalogs, documenting the discoveries of binary and multiple cluster systems to date, utilizing several methods and techniques, and the 3D (or 6D) phase-space information. These catalogs encompass diverse datasets, reflecting the evolving approaches to identifying and analyzing these systems. Recent studies have focused on creating a homogeneous sample of Galactic star clusters with consistent membership criteria. Notably, many of these endeavors rely on data from the Gaia mission \citep{Gaia18}, which offers high-precision photometric and astrometric measurements, facilitating a more accurate identification and analysis of these systems \citep[see e.g.,][]{Cantat+20, Hunt2023, Hunt2024}. 

\begin{table} 
   \caption{Compilation of multiple systems detection found in the literature}
   \label{tab:ref}
   \centering
   \resizebox{\columnwidth}{!}{ 
   \begin{tabular}{lcc}
   \hline
   Authors & N$^{\circ}$ binary systems & N$^{\circ}$ groups \\
   &  candidates & candidates \\
   \hline
   \citet{Pavlov89} & -- & 5  \\
   \citet{Subra95} & 18 & --  \\
   \citet{delaFuente09} & 43 & --  \\
   \citet{Conrad17} & 14 & 5  \\
   \citet{Soubiran19} & 8 & 4  \\
   \citet{Liu19} & 39 & 16  \\
   \citet{Zhong19} & 1 & --  \\
   \citet{PieckaPaunzen21} & 50 & 10 \\
   \citet{Casado21a} & 11 & 11 \\
   \citet{Casado21b} & 1 & -- \\
   \citet{Angelo+22} & 5 & 2 \\
   \citet{Song22} & 14  & -- \\
   \hline
   \end{tabular}
   }
\end{table}

In this study, we independently identified new star cluster pairs and groups by employing a method that considers the tidal forces exerted on each cluster by its closest neighboring cluster. This approach allows us to discern potential gravitational interactions and connections between clusters, which may not be apparent through traditional methods, especially in populated regions. By focusing on these tidal influences, we can more accurately identify clusters that are likely part of binary or multiple systems.

The structure of this paper is as follows: Section \ref{sec:sample} describes the sample used in this study. Next, in Section \ref{sec:candidates}, we present the newly identified candidate binary and multiple systems. Finally, Section \ref{sec:conclusions} summarises our conclusions.

\section{Sample of star clusters}
\label{sec:sample}

\subsection{The sample}

Our study is based on the catalogs created by \citet[][HR23 hereafter]{Hunt2023} and \citet[][HR24 hereafter]{Hunt2024}. In these articles, the authors undertook an extensive all-sky search for open clusters using data from the Gaia DR3\footnote{\url{https://www.cosmos.esa.int/web/gaia/dr3}} release \citep{Gaia:dr3}. This has resulted in a homogeneous catalog containing 7167 clusters, of which 4782 are known from the literature and 2387 are newly identified candidates. To recover and detect these clusters, they employed the HDBSCAN (Hierarchical Density-Based Spatial Clustering of Applications with Noise) algorithm, which is adept at identifying clusters of varying densities and separating them from noise. In addition to cluster identification, these studies provided detailed inferences on a range of fundamental parameters for each cluster. These include basic astrometric parameters such as proper motions and parallaxes, and astrophysical properties like ages, extinctions, distances, and photometric masses. The authors also distinguished between bound and unbound clusters, offering insights into the dynamic states of these systems. They found that only 79\% of the clusters in their catalog were consistent with being bound open clusters (OCs), while the rest were identified as moving groups (MGs). This rich dataset serves as a crucial foundation for our analysis, enabling a thorough exploration of the characteristics and dynamics of previously both known and newly discovered star clusters. 

For our purpose, we made use of the physical properties derived for the entire sample of Hunt \& Reffert catalogs \citepalias{Hunt2023, Hunt2024}, specifically focusing on total mass, size, and age. Briefly, the total cluster mass $M$, was estimated by determining the photometric masses of member stars using PARSEC isochrone fits in the G-band \citep{Bressan:2012}, followed by corrections for selection effects and unresolved binaries. Then, they fitted a mass function to each cluster and integrated it to estimate the total mass, adopting a \citet{Kroupa:2001} initial mass function and employing the \texttt{imf} Python package\footnote{\url{https://github.com/keflavich/imf}}. The size $R_{50}$ used in this paper corresponds to the radius that encloses 50\% of the cluster members, while the ages of the star clusters were determined by the authors using a neural network.

\subsection{Systems identification}

We first used the 3D cartesian heliocentric Galactic coordinates of the star clusters to identify the corresponding closest neighbor to each one. In this process we used the \texttt{Sklearn.neighbors.NearestNeighbors}\footnote{\url{https://scikit-learn.org/stable/}} Python package to efficiently determine the nearest neighbor to each cluster. Those star clusters with reciprocity were therefore identified as closest neighbors. As a second step, we estimated the tidal force exerted on each cluster by considering only the influence of its closest neighbor. The tidal force estimation was calculated using the following tidal factor: 

\begin{equation*}
\centering
{\it TF} =  \frac{d^3}{M_{b} R_{50}} 
\end{equation*}

\noindent where $d$  is the distance to the closest neighbor in parsecs; $M_{b}$ is the total mass, in solar masses, of the neighbor cluster; and $R_{50}$ is the radius, in parsecs, containing 50\% of members within the tidal radius. These latter two parameters are given by \citetalias{Hunt2024}.  It must be noticed that {\it TF} is the reciprocal of the tidal force, therefore a low {\it TF} value corresponds to a high tidal force and vice versa. This procedure classifies the clusters as belonging to {\it Pairs} or {\it Groups}. The remaining ones were considered as a mixture of {\it Singles} and unclassified ({\it Unclass}). 

To define our selection and boundaries criteria, we cross-referenced the \citetalias{Hunt2024} catalog with various other catalogs of double and multiple cluster systems, as listed in Table \ref{tab:ref}. With this new compilation, we analyzed the distributions of 3D distances to the closest neighbor, {\it TF}, and sizes for each star cluster category. Fig.\ref{fig:dist} present these distributions and reveals that most star clusters considered part of a system ({\it Pairs} or {\it Groups}) predominantly have distances less than $\sim$ 50-100 pc and {\it TF} values below $\sim$ 250, without any notable differences among size distributions.  Taking into account these distributions, we categorized the star clusters according to the following classification rules:

\begin{itemize}
    \item {\it Pairs} (P): These are sets of two close ($<50$\,pc) neighbor clusters with at least one of them having a low {\it TF} value ($<200$), without any third nearby cluster.
    \item {\it Groups} (G): These are sets of three or more clusters. They were identified originally as pairs, but with at least one additional close cluster ($<50$\,pc) with low {\it TF} value ($<200$). 
    \item {\it Singles} (S): These are star clusters located beyond $100$\,pc their nearest neighbor.
    \item {\it Unclass}: Remaining clusters not included in previous cases.
\end{itemize}

\begin{figure}
\centering
\includegraphics[width=0.8\columnwidth]{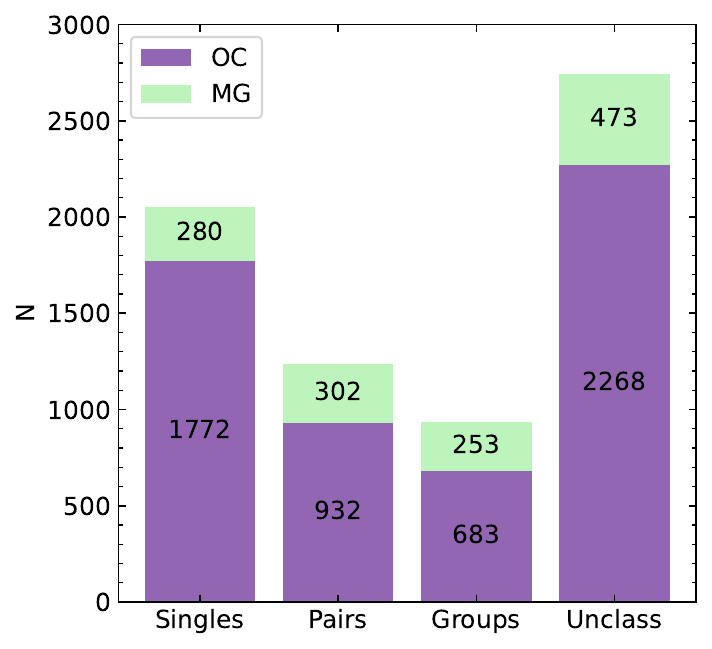}
\caption{OCs (violet) and MGs (lightgreen) distribution into cluster systems based on our {\it TF} classification method.}
\label{fig:N}
\end{figure}

We analyzed the newly obtained distributions of closest neighbor distances, {\it TF}, and sizes for each star cluster category, following our new classification criteria (Fig.\ref{fig:dist2}). We identified a total of 2170 candidate star clusters (OCs and MGs) within pairs or groups. A summary of the stellar systems found is given in Fig. \ref{fig:N}. For each category ({\it Singles, Pairs, Groups, Unclass}) we distinguish among the OCs and MGs.

\begin{figure}
\centering
\includegraphics[width=0.8\columnwidth]{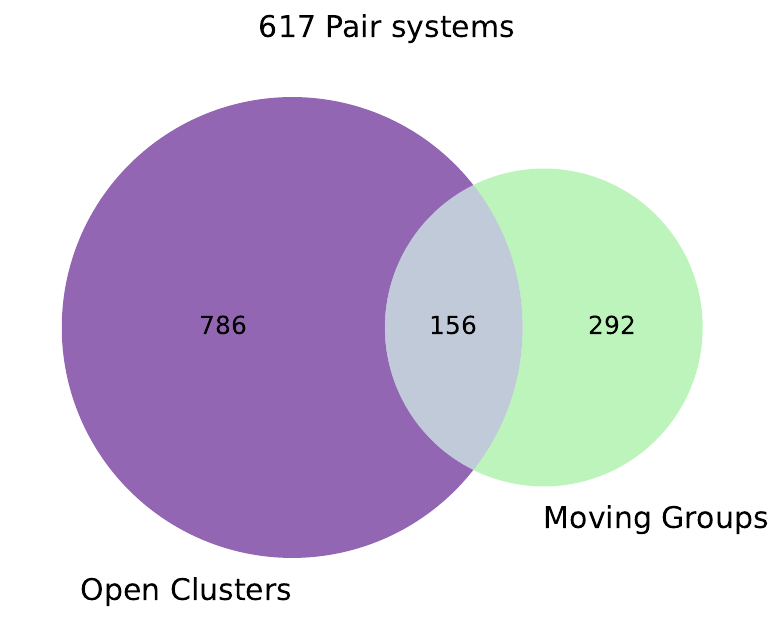}
\caption{Number of star clusters
of pair systems composed of OCs (violet), MGs (light green) or a combination of both.}
\label{fig:venn_pairs}
\end{figure}

\begin{figure*}
\centering
\includegraphics[width=\textwidth]{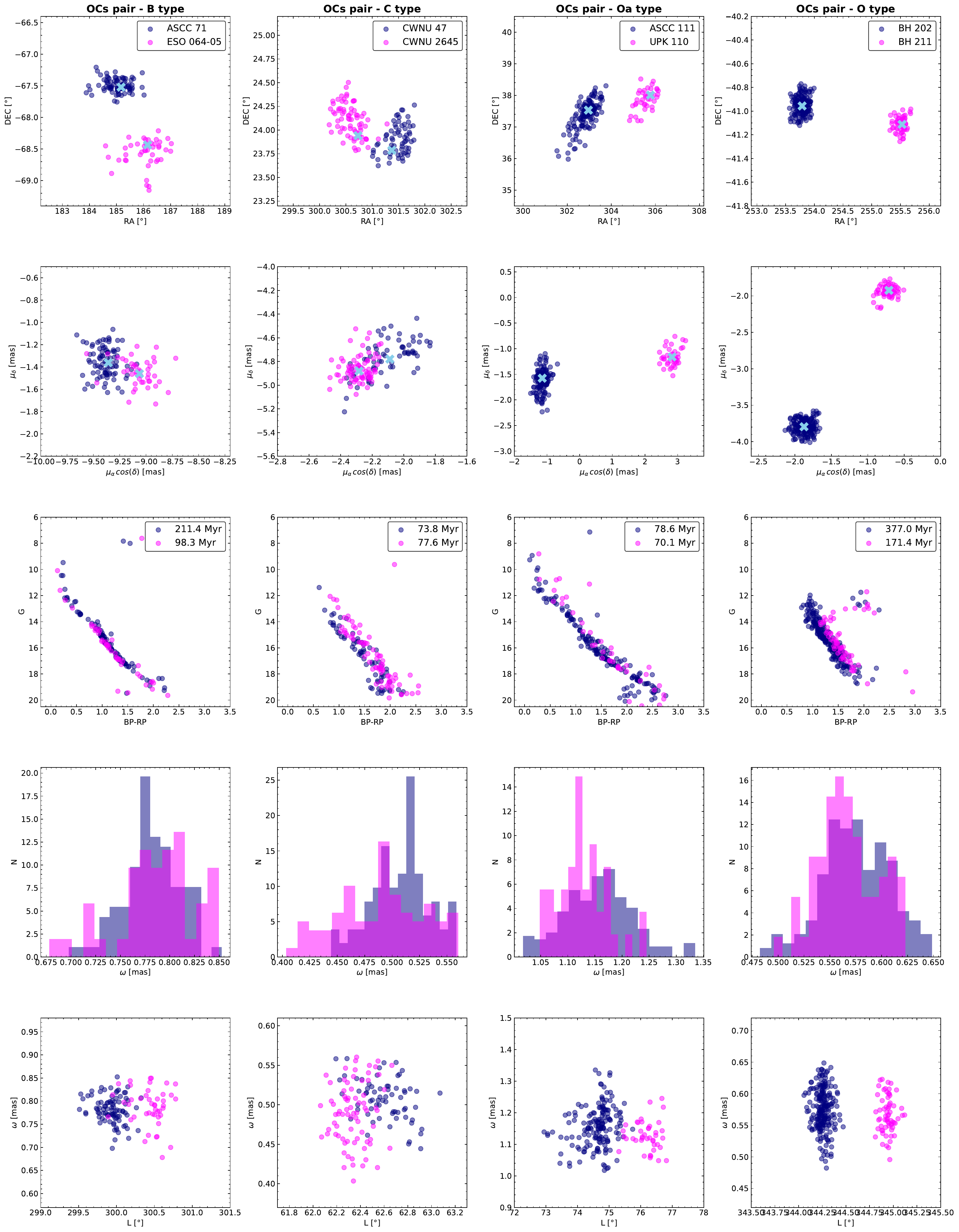}
\caption{Distributions in a multidimensional space (positions, proper motions, and parallaxes) of member stars with at least $70\%$ of membership probability, of different types of selected pairs. All the clusters are identified as OCs. The light blue crosses mark the central cluster coordinates or clusters PMs.}
\label{fig:OC_bin}
\end{figure*}

\section{New binary and multiple system candidates}\label{sec:candidates}

Based on our identification and classification criteria, we compiled new catalogs of binary and multiple cluster candidates. We included the fundamental and relevant parameters for a total of 2170 OCs and MGs, as determined by the \citetalias{Hunt2023, Hunt2024} catalogs and our analysis. The identification process involved careful analysis of cluster properties such as position, parallax, proper motions (PMs), and color-magnitude diagrams (CMDs) published on these catalogs. Our compilation and study have been conducted homogeneously, as we applied the same procedures and criteria across all datasets. This consistency ensures that our results are directly comparable with previous studies and that any newly identified cluster systems are reliably classified.

\subsection{Pair star clusters}

Out of the 617 binary system candidates identified using the {\it TF} criteria, 52 {\it Pairs} have been previously cataloged as part of binary or multiple cluster systems. Notably, our method successfully rediscovered 7 binary systems that were already documented in earlier catalogs, validating the reliability of our approach. The remaining 45 {\it Pairs} are analyseed and discussed in section \ref{sec:cat}. This cross-verification not only reinforces the accuracy of our methodology but also highlights its effectiveness in identifying genuine binary cluster systems. In addition to confirming known systems, our analysis revealed 610 new candidate pair clusters that have not been previously identified. For each cluster, we carefully selected stars with membership probability greater than 70\% to ensure the accuracy and reliability of our identifications. This threshold was chosen to minimize contamination from field stars and to provide a clearer understanding of the clusters true members. Fig. \ref{fig:venn_pairs} shows, in a Venn diagram, the number of pair systems in the different compositions, where both members are OCs, MGs, or a combination of both.

To evaluate the potential physical interactions within these candidate binary systems, we analyzed their member stars distributions in a multidimensional space, which includes coordinates, parallax, and PMs, as well as their corresponding CMDs. This comprehensive assessment allowed us to consider multiple aspects of the clusters' physical properties simultaneously, to better understand whether they constitute physically interacting systems. Based on these analyzes, we classified the pair systems (P) candidates into the following categories: 
\begin{itemize}
    \item Genetic pairs or binaries (B): These are clusters that form simultaneously, sharing common properties such as distance (parallax), kinematics (compatibility in PMs), and age. They likely originated from the same molecular cloud and have similar evolutionary histories.
    \item Tidal capture or resonant trapping pairs (C): These systems consist of clusters that occupy a limited volume of space and share common kinematics but do not necessarily have similar ages. The shared kinematics suggest that these clusters are gravitationally influenced by each other, possibly due to tidal capture or resonant trapping mechanisms. This indicates a dynamic interaction without a common origin in time.
    \item Optical pairs (O): These clusters appear close together in the sky but are not gravitationally bound. They result from gravitational (hyperbolic) encounters \citep{delaFuente09} and may not share common kinematics. Optical binaries are further subdivided into two types: Clusters that occupy a limited space and have similar ages, suggesting they pass near each other but are not dynamically associated (Oa). Clusters that occupy a limited volume of space but do not have similar ages indicate a chance alignment in the sky without any significant gravitational interaction (O).
\end{itemize}

\begin{figure}
\centering
\includegraphics[width=0.85\columnwidth]{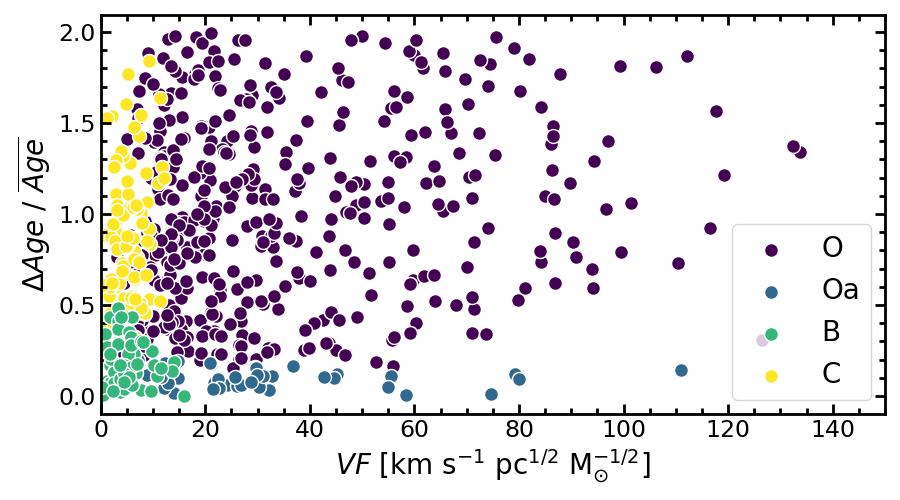}
\caption{Ages and virial state for star clusters in pair systems.}
\label{fig:ages_virial}
\end{figure}

An example of each category is given in Fig. \ref{fig:OC_bin} for OC pairs. 
This classification framework allows us to distinguish between various types of binary systems, ranging from those with strong physical associations to merely coincidental alignments. The following physical features were used to characterize those clusters in pairs: a) the relative age difference ($\Delta Age / \overline{Age}$); and b) a virial factor ($VF$) defined as:
\begin{equation*}
    VF = v_T \sqrt{\frac{\overline{R_{50}}}{\overline{M}}}
\end{equation*}
\noindent where $\overline{R_{50}}$, and $\overline{M}$ are the mean size (in pc) and the mean mass (in solar masses), respectively, of both clusters; and $v_T = 4.74\,\Delta\mu\,D_{50}$ is the relative tangential velocity between the clusters (in km/s). The corresponding distribution of star clusters in pairs is presented in Fig.~\ref{fig:ages_virial}. This plot reveals that the previously suggested categories are easily separated using the limiting values $VF\sim10$\,km\,s$^{-1}$\,pc$^{1/2}$\,M$_{\odot}^{-1/2}$ and $\Delta Age~/~\overline{Age} \sim 0.2 - 0.5$. \\

Table \ref{tab:bin} summarizes the relevant properties of our newly detected pair systems. The column {\it kind} distinguishes between OCs (o) and MGs (m). The mass ratio is given as $M_A / M_B$ and $\Delta$Age$_{50}$ represents the median age pair difference. We consider that differences in Age$_{50}$ below 10\%, in logarithmic scale, are indicating coeval clusters. PM$_{compat}$ denotes the compatibility of the clusters' PMs within 3$\sigma$, and BCO refers to the pair classification. 

\begin{figure}
\centering
\includegraphics[width=0.8\columnwidth]{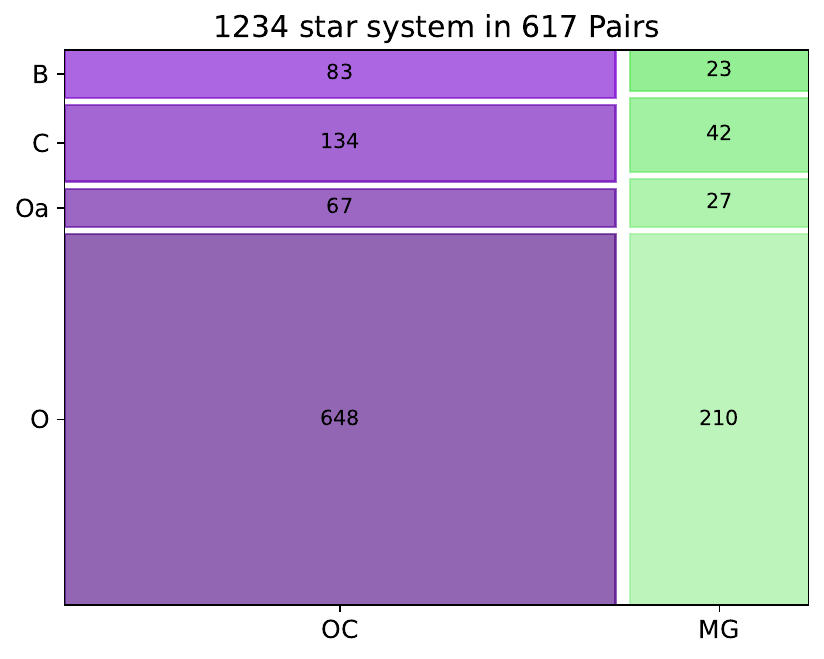}
\caption{Distribution of OCs in pair systems, across each type and category. }
\label{fig:mosaico}
\end{figure}

Fig. \ref{fig:mosaico} summarize the number of stellar systems in each pair category, distinguishing among OCs and MGs. As can be observed and expected, genetic binary systems are the minority. 
By examining these different categories, we can better understand the formation and evolution of these stellar systems.\\

\begin{table*}[!ht]
   \caption{Binary system candidates. Col. (1): Pair IDs. Col. (2): Cluster names. Col. (3): HR24 Internal ID. Cols. (4), (5): J2016.0 {\bf cluster coordinates}. Col. (6): Object type estimated by HR24. Col. (7): Tidal factor in $pc^2/M_{\odot}$. Col. (8): Median age in Myr, estimated by HR23. Col. (9): The mass ratio between the clusters in the pair. Col. (10): Median age differences of the pair clusters. Col. (11): Compatibility in proper motions. Col (12) Pair classification. This table is fully available in the electronic edition; a portion is shown here for illustrative purposes regarding format and content. }
  \label{tab:bin}
  \centering
 \begin{tabular}{clrrcccccccc}
\hline
\hline
\vspace{1pt}
 Pair & Cluster & ID$_{HR24}$ & RA  & DEC & kind & {\it TF} & Age$_{50}$ & $M_A/M_B$  & $\Delta$Age$_{50}$ & PM$_{compat}$ & BCO \\
   &  &  & [deg] & [deg] &  & [$pc^2/M_{\odot}$] & [Myr] &  & [Myr] &   & \\
\hline
 P1 & Andrews-Lindsay\_5 & 118 & 281.078 & -4.930 & o & 1.28 & 108.48 & 2.84 & 62.98 & False & O \\
 \vspace{4pt}
  & HSC\_279 & 2006 & 281.276 & -4.859 & o & 0.36 & 45.50 & & & & \\
 P2 & BDSB\_30 & 125 & 333.663 & 61.444 & o & 7.15 & 8.30 & 2.25 & 5.04 & True & C \\
 \vspace{4pt}
  & OC\_0181 & 4941 & 334.027 & 60.912 & o & 1.82 & 3.26 & & & & \\
 P3 & Theia\_1806 & 5750 & 85.433 & 27.500 & o & 96.97 & 187.65 & 2.20 & 162.63 & False & O \\
 \vspace{4pt}
  & COIN-Gaia\_21 & 310 & 84.755 & 28.437 & o & 192.61 & 25.02 & & & & \\
 P4 & CWNU\_1229 & 626 & 20.904 & 61.836 & o & 98.42 & 17.68 & 1.46 & 68.82 & True & C \\
 \vspace{4pt}
  & HSC\_1031 & 2657 & 24.569 & 61.566 & o & 30.64 & 86.50 & & & & \\
 P5 & CWNU\_133 & 387 & 49.923 & 52.687 & o & 18.53 &  108.14 & 0.73 & 66.21 & True & C \\
  & Czernik\_15 & 1366 & 50.818 & 52.218 & o & 13.38 &  41.93 & & & & \\
\hline
\end{tabular}
\end{table*}

\begin{figure}
\centering
\includegraphics[width=0.7\columnwidth]{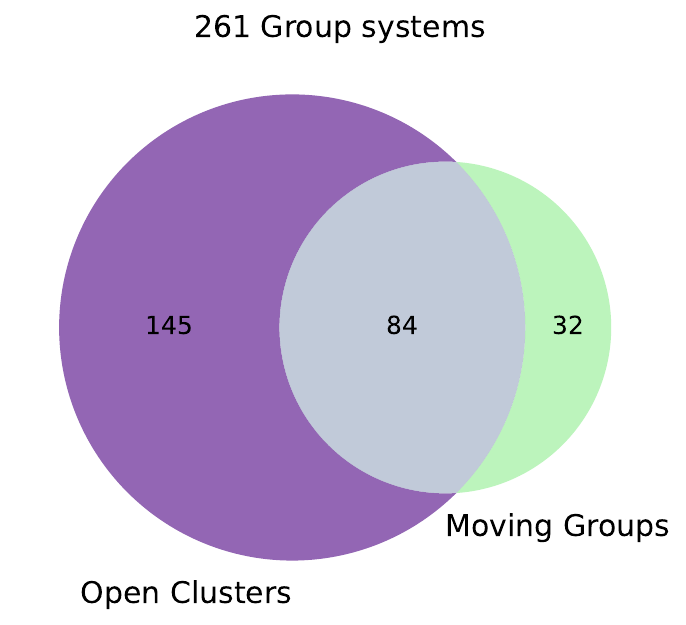}
\caption{Composition of new multiple cluster systems. The diagram shows the number of systems formed entirely of OCs (violet), MGs (light green), or a combination of both.}
\label{fig:venn_groups}
\end{figure}
\begin{figure}
\centering
\includegraphics[width=0.8\columnwidth]{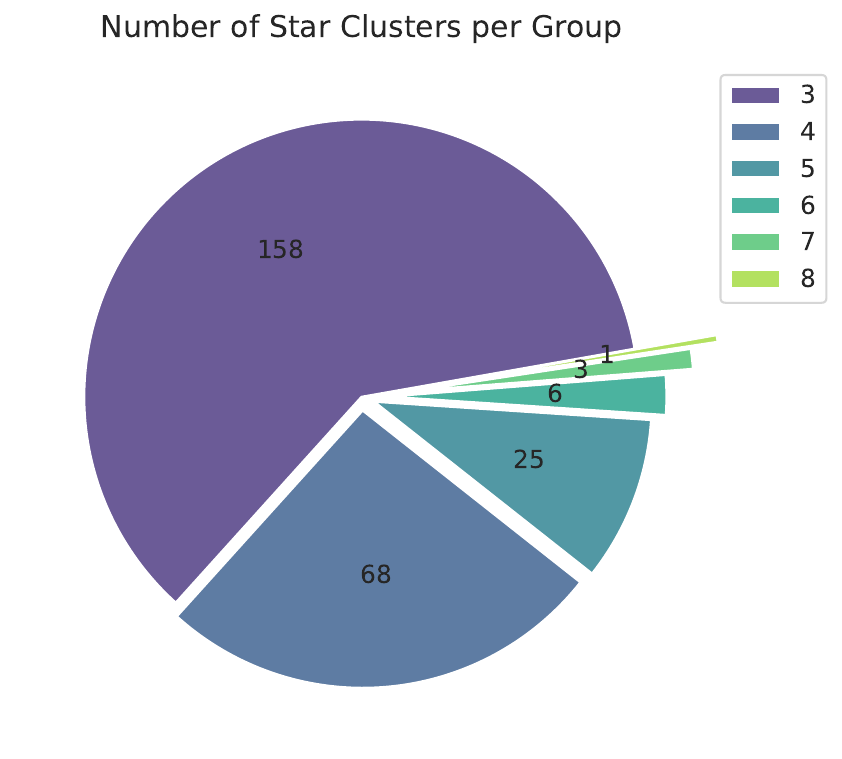}
\caption{The pie chart shows the number of star clusters in each group. Most of the systems consist of groups with 3-star clusters.}
\label{fig:pie_groups}
\end{figure}

\subsection{Group star cluster candidates}

\begin{figure*}
\centering
\includegraphics[width=\textwidth]{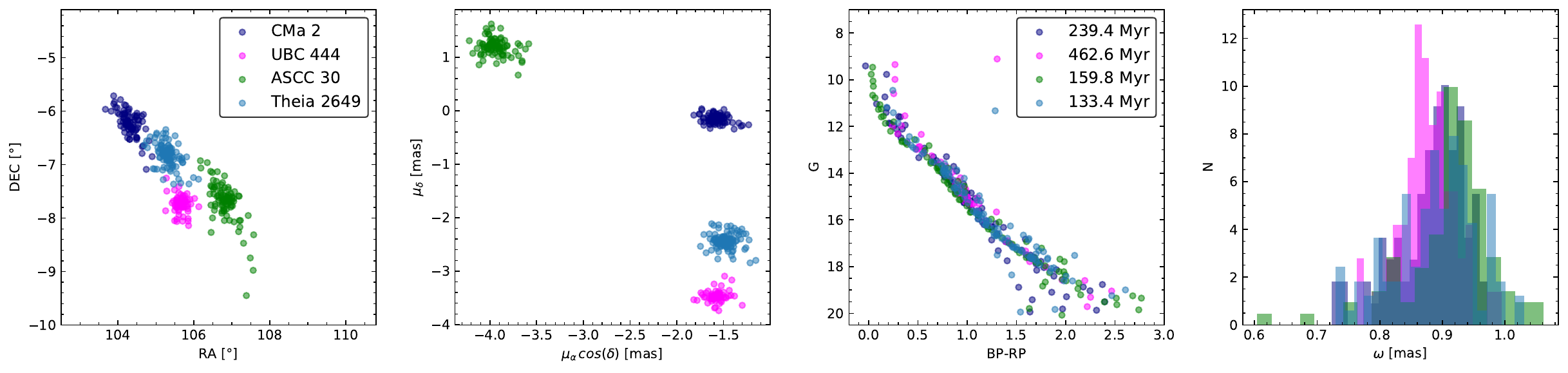}
\caption{Distributions in a multidimensional space (positions, proper motions and parallaxes) of member stars with at least $70\%$ of membership probability, of a selected group. All the clusters involved are OCs.}
\label{fig:OC_gr}
\end{figure*}

The {\it TF} criteria  identified 936 star clusters, forming 261 multiple systems, i.e. groups with three or more components. Table \ref{tab:grp} summarizes the relevant properties of these newly detected multiple systems.  Fig. \ref{fig:venn_groups} illustrates the configurations of these groups, which consist entirely of OCs, MGs, or a combination of both.
Fig. \ref{fig:pie_groups} presents the number of star clusters, composed of OC and/or MG, within each system, as a function of the number of members. Most systems are made up of groups with three star clusters, followed by groups with four and five members. There is only one system with eight star clusters.
Applying the same procedure as for binary systems, we selected stars for each cluster with a membership probability greater than 70\%. We then analyzed their distributions across spatial coordinates, proper motions, and parallax planes, as well as their CMDs to assess the systems' physical properties. An example of our analysis, along with the results for a selected multiple cluster system, is given in Fig. \ref{fig:OC_gr}.

 The cluster members within each group exhibit different degrees of association with their respective group members. Some share common PMs and/or ages, while others are coincidentally aligned and share the same spatial volume. To quantify these characteristics, we computed their parameters $\sigma Age~/~\overline{Age}$ and $VF$. Then, we adopted similar limit values to those obtained from Fig.~\ref{fig:ages_virial} for clusters in pairs, and we estimated that about 7\% of the identified clusters in groups could be considered part of coeval systems.\\ 

\begin{table*}[!ht]
   \caption{Multiple system candidates. Col. (1): Group IDs. Col. (2): Cluster names. Col. (3): HR24 Internal ID. Cols. (4), (5): J2016.0 cluster coordinates. Col. (6): Object type estimated by HR24. Col. (7): Tidal factor in $pc^2/M_{\odot}$. Col. (8): Median age in Myr, estimated by HR23. Col. (9): Previous references. Col. (10): Number of clusters in each group. This table is fully available in the electronic edition; a portion is shown here for illustrative purposes regarding format and content.}
  \label{tab:grp}
  \centering
 \begin{tabular}{clcrrccccc}
\hline
\hline
\vspace{1pt}
   Group & Cluster & ID$_{HR24}$ & RA &  DEC & kind  &   {\it TF} &  Age$_{50}$ &  REF & N$_{cls}$ \\
    &  &  & [deg] & [deg] &  & [$pc^2/M_{\odot}$] & [Myr] &  &  \\
\hline
G1 & Alessi\,3 & 69 & 109.187  & -46.680  & o &  133.73  &  625.55    &  $a$  & 4 \\
& CWNU\,536 & 558  & 105.662 & -45.679  & m &  41.70 &  145.54   & ...  &  \\
& CWNU\,1031 & 571  & 116.141 & -43.504  & m &  34.12 &  144.45   & ... &  \\
\vspace{4pt}
& CWNU\,1265  &  634 & 113.288  & -49.138 & m  &  21.75 & 1160.09   & ... & \\ 

G2 & Alessi\,33  & 84 &  105.710  & -26.365 & o &  136.09  &   15.69 & ...  & 3 \\
& HSC\,1894  & 3315 & 103.267  & -26.974 &  m  &  29.94 &   20.88  & ... & \\
\vspace{4pt}
& HSC\_1913 & 3330 & 106.266 & -27.796 & o  &  51.82 &  11.72  & ... & \\
G3 & Alessi\,36  & 86 & 106.565 & -37.596 & o &  36.70 &   28.90  & ...  & 3 \\
  & Collinder\,135  & 1323 &  109.406  & -36.921 & o &  36.19 &   29.83  & $a,b,c,d,e,f$ & \\
\vspace{4pt}
  & HSC\,1949  & 3360 & 107.218  & -31.567 & m  & 72.78 &  424.70  & ... & \\
G4 & Alessi\_62 & 94 & 283.963 & 21.607  & o & 30.62 & 425.43 &  $a,b$ & 3 \\ 
  & HSC\_429 & 2128 & 283.436 &  21.117  & m &  1.52 &  269.20 & ... & \\
  \vspace{2pt}
  & UBC\_26  & 6080 & 285.426 & 22.113  & o & 1.48 &   25.16 & ... & \\
\hline
\end{tabular} \\
\tablefoot{$^a$\citet{Liu19}; $^b$\citet{PieckaPaunzen21};$^c$\citet{Conrad17};$^d$\citet{Soubiran19};$^e$\citet{Casado21b};$^f$\citet{Song22} }
\end{table*}

\begin{figure*}
\centering
\includegraphics[width=\textwidth]{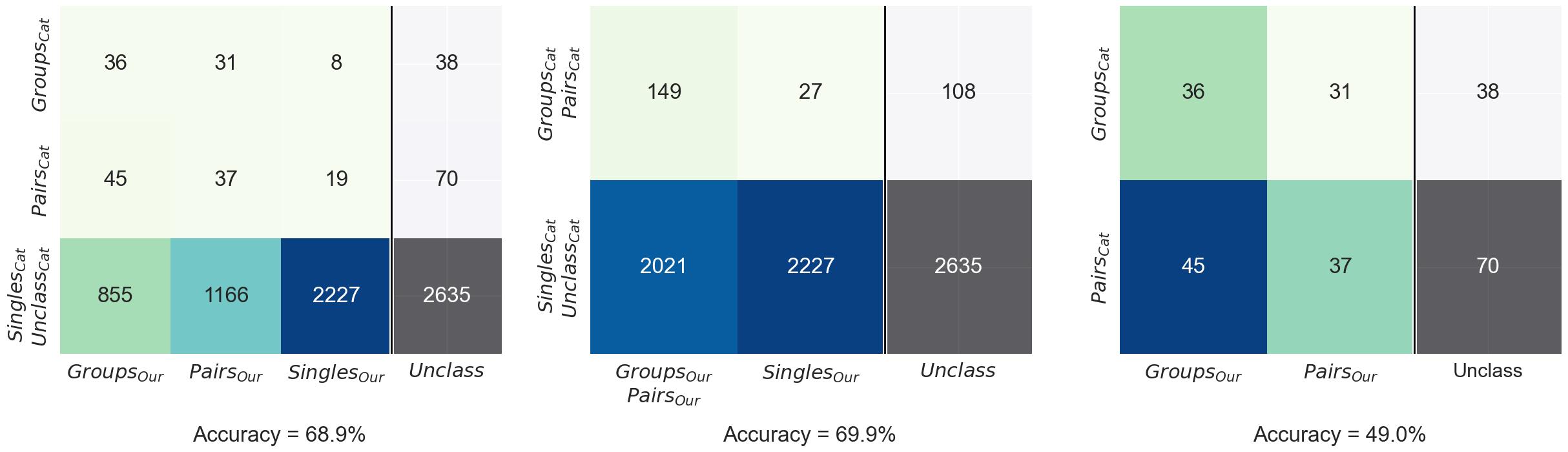}
\caption{Confusion matrices comparing star cluster classifications from previous studies ("$Cat$" items) with our systematic method, which uses the tidal factor, distances, and pair identification (see Sect. \ref{sec:sample}). The left matrix categorizes star clusters into {\it Groups}, {\it Pairs}, {\it Singles}, and unclassified ({\it Unclass}). The central matrix combines the {\it Groups} and {\it Pairs} categories, while the right matrix considers only the {\it Groups} and {\it Pairs} classifications. The corresponding accuracy for each matrix is also provided.
}
\label{fig:matrices}
\end{figure*}

\subsection{Analysis of cataloged binary and multiple clusters}
\label{sec:cat}

The cross-correlation of \citetalias{Hunt2024} catalog with other catalogs of double and multiple clusters systems, not only helped us define our selection criteria but also enabled us to validate and expand existing data. That is, verifying the presence of known binary and multiple systems while also identifying new candidates. To compare classification criteria, we constructed confusion matrices, as shown in Fig.~\ref{fig:matrices}. The results demonstrate an overall accuracy of about 70\% (left and central matrices) in distinguishing between the different categories and when combining {\it Singles} and {\it Unclass} star clusters according to our criteria. On the other hand, the specific discrimination between {\it groups} and {\it pairs} shows an accuracy of approximately 50\% (right matrix). In particular, our classification rules reveal: 
\begin{itemize}
    \item 2021 star clusters as {\it Pairs} or {\it Groups} that were previously unclassified or considered as {\it Singles}.
    \item 45 star clusters being part of {\it Groups} that were previously considered part of {\it Pairs}.
    \item 27 star clusters identified previously as being part of $Groups$ or $Pairs$, are now identified as {\it Singles}.
    \item 31 star clusters considered previously as being part of {\it Groups} are now classified as {\it Pairs}.
\end{itemize}

\section{Conclusions}               
\label{sec:conclusions}

In this work, we employed a new method to identify star clusters in the Milky Way that are part of double or multiple systems. For this purpose, we used the \citet{Hunt2023, Hunt2024} database, which comprises 7167 star clusters. By estimating the tidal force through the tidal factor ({\it TF}), we identified 1234 star clusters forming 617 paired systems. These systems consist of open clusters, moving groups, or a combination of both. Taking into account stars membership probabilities within each cluster, we further classified the systems into three categories: Binaries (B), Capture pairs (C), and Optical pairs (O\,/\,Oa). This classification was determined using proper motion distributions, cluster ages, and color-magnitude diagrams.  Additionally, we identified 936 star clusters forming 261 groups, each with three or more members. Our sample increases the currently known sample of double and multiple systems in the Milky Way by approximately 2.7 and 4.5 times, respectively. Our method offers a more accurate approach for identifying clusters within the same spatial volume that exhibit significant tidal interactions with their neighbors. \\

\begin{acknowledgements}
We gratefully acknowledge financial support from the Argentinian institutions: Consejo Nacional de Investigaciones Cient\'ificas y T\'ecnicas (CONICET; PIP-2022-11220210100064CO and PIP-2022-11220210100714CO), Agencia Nacional de Promoción de la Investigación, el Desarrollo Tecnológico y la Innovación (PICT-2020-3690), Secretaría de Ciencia y Tecnología de la Universidad Nacional de Córdoba (SECYT-UNC, Res. 258/53) and from the Universidad Nacional de La Plata (project: 11/G182).
\end{acknowledgements}


\bibliographystyle{aa} 
\bibliography{paper1} 

\end{document}